\documentclass[aps,preprint,superscriptaddress,floatfix,pra]{revtex4-1}
\usepackage[utf8]{inputenc}
\usepackage{amsmath}
\usepackage{amssymb}
\usepackage[export]{adjustbox}
\usepackage[colorlinks=true,citecolor=blue,linkcolor=blue,urlcolor=blue]{hyperref}
\usepackage[capitalise]{cleveref}

\usepackage[english]{babel}
\usepackage{comment}
\usepackage{braket}
\usepackage[toc,page]{appendix}
\usepackage{graphicx}
\graphicspath{ {Figures/} }

\usepackage[makeroom]{cancel}
\usepackage{textcomp}

\usepackage[usestackEOL]{stackengine}
\usepackage{wrapfig}
\usepackage{mathtools}
\usepackage{physics}
\usepackage{multirow}
\usepackage{float}
\usepackage{xcolor}

\usepackage[twocolumn,textwidth=17.2cm,columnsep=0.7cm,top=15mm,
bottom=15mm]{geometry}

\makeatletter
\renewcommand{\fnum@figure}{FIG. \thefigure}
\makeatother
\usepackage{relsize}

\usepackage{enumitem}

\usepackage[nodisplayskipstretch]{setspace}
\usepackage{titlesec} 
\titlespacing*{\section}{0cm}{10pt}{10pt}

\begin{document}

\title{Level crossings and Fisher information of two hard-core bosons in a two-dimensional trap} 

\author{Dimitris Saraidaris}
\email[]{dimitris.saraidaris@gmail.com}
\affiliation{Department of Physics, University of Athens, 15784 Athens, Greece }

\author{Ioannis Mitrakos}
\email[]{Johnmit95@yahoo.gr}
\affiliation{Department of Physics, University of Athens, 15784 Athens, Greece }

\author{Ioannis Brouzos}
\email[]{ibrouzos@phys.uoa.gr}
\affiliation{Department of Physics, University of Athens, 15784 Athens, Greece }

\author{Fotios Diakonos}
\email[]{fdiakono@phys.uoa.gr}
\affiliation{Department of Physics, University of Athens, 15784 Athens, Greece }

\begin{abstract}
We investigate a  two-body quantum system with hard-core interaction potential in a two-dimensional harmonic trap. We provide the exact analytical solution of the problem. The energy spectrum of this system as a function of the range of the interaction reveals level crossings which can be explained examining the potential and kinetic part of the energy, as well as the localization and oscillatory properties of the wavefunctions quantified by the Fisher information. The latter is sensitive to the wavefunction localization between nodes promoting this property as a resource of quantum correlations in interacting particle systems. 
\end{abstract}

\date{\today}


\maketitle 

\section{INTRODUCTION}

Correlated few body-systems are of fundamental importance for several fields of physics~\cite{review_blume_fewbody,review_greene_fewbody,review_nuclear_fewbody}. Understanding the correlations on a two-body level is essential to build up models for many-body systems. More particularly, for  cold atom physics, the two seminal works of Olshanii~\cite{olshanii_1998}, who derived a pseudopotential for quasi-1D systems, and Busch~\cite{busch_1998}, who derived analytical solutions for the two-body problem in 1-, 2- and 3-D traps, have determined the field.

In the case of 1D few-body systems, a unique phenomenon appears when we assume that the zero-range interaction potential tends to infinity. Girardeau~\cite{girardeau} has shown that bosons can be mapped to free fermions and form the celebrated Tonks gas and the 'fermionization' of bosons. The map is valid for all local properties (energy, density) and for any number of atoms from two to many. This seminal work together with Lieb-Liniger model~\cite{liebliniger} with exact solution for arbitrary interaction in the continuum, have given rise to a lot of studies in 1D~\cite{1d1,1d2}.

When we increase the dimensionality, bosonic systems are more subtle since the theorem of Girardeau for the mapping does not apply. Nevertheless, for two- and few-body systems previous works~\cite{bruno,bruno2,b3,b4} have shown that still there is a tendency for two bosonic particles to avoid being on the same position as the repulsive interaction potential increases as a reminecent of the 'fermionization' effect. This is shown in the density profile and at the level of correlation functions.

A different point of view in the density profile is given by Fisher information, a local information-theoretic quantity, very sensitive to local rearrangements of the density \cite{r1,r2,romera}. In central potentials, relevant studies have shown the connection between Quantum Fisher Information (QFI), the localization of the density and the oscillatory behaviour of the wavefunction for single-particle problems. Beside this, the notion of Fisher information for quantum systems has a broader application. It has been shown for instance, that Fisher information is connected to energy level crossings~\cite{levelcrossing_fisher} for a two-level system. 

In this work, we examine the two-body problem with a hard-core potential in a 2D parabolic trap. In this trap the centre-of-mass  decouples from the relative-coordinates, thus we focus on the latter. We derive first the exact analytical solutions of the corresponding Schr\"{o}dinger problem in terms of hypergeometric functions~\cite{hyper,Hass,confined} applying on top the hard-core interaction condition. Then, we explore the behaviour of the energy spectrum as a function of the range of the hard-core interaction. We find that the calculated spectrum has the unique property of level-crossings. We explain the dependence of the energy levels and the crossings on the range of the hard-core interaction, separating the energy into parts (potential and kinetic) and considering the impact of each part on the corresponding modulations on the wavefunction. These modulations, as well as the crossings, are finally connected to Fisher information, which quantifies the qualitative  observations and provides a link to quantum correlations.

\section{SOLUTION FOR HARD-CORE 2D TRAPPED PARTICLES}

This section is organized in four subsections. In subsection \ref{hamilt} we first pose the problem of two hard-core interacting particles in a 2D harmonic trap and then we follow a solution strategy in three steps presented in subsections \ref{seph}, \ref{relwf} and \ref{2bcondition}.

\subsection{Hamiltonian}\label{hamilt}
The non-interacting part of the Hamiltonian for two identical particles in a 2D harmonic potential reads:

\begin{equation}
\mathcal{H}_0=\frac{\mathbf{p}_{r_1} ^2}{2m} + \frac{\mathbf{p}_{r_2} ^2}{2m} + \frac{1}{2}m\omega^2\mathbf{r}_1^2+\frac{1}{2}m\omega^2\mathbf{r}_2^2
\end{equation}
\noindent where $\mathbf{r}_1,\mathbf{r}_2$ are the 2-D particles' position vectors. The full Hamiltonian includes also the interacting potential $H=H_0+V(|\mathbf{r}_1-\mathbf{r}_2|)$. We assume an interaction potential of a hard-core nature at distances shorter than a parameter $r_0$, which corresponds to the range of the hard-core interaction: 
\begin{equation}
\label{eq:2}
       V(|\mathbf{r}_1-\mathbf{r}_2|)=
        \left\{ \begin{array}{ll}
            0, & |\mathbf{r}_1-\mathbf{r}_2|> r_0 \\
            \infty, & |\mathbf{r}_1-\mathbf{r}_2|\leq r_0\\
        \end{array} \right.
    \end{equation}

\noindent Effectively in 2D this corresponds to a classical picture of hard-core disks, where $r_0$ is the diameter.

It is straightforward to solve the problem analytically: after separating the Hamiltonian for the non-interacting part $H_0$ in centre-of-mass (CM) and relative coordinates in \ref{seph}, we solve in \ref{relwf} the Schr\"{o}dinger problem in relative coordinates, using an ansatz for the wave function in terms of hypergeometric functions, suitably adapted to have the correct asymptotic behaviour. Then, in \ref{2bcondition} we determine the energy spectrum imposing the appropriate boundary conditions for the relative coordinates, as dictated by the two-body interaction potential.

\subsection{Separation of the Hamiltonian} \label{seph}

Due the parabolic nature of the external potential, the non-interacting part of the Hamiltonian $H_0$ decouples into two parts depending on CM:
\begin{equation}
\mathcal{H}_{0,\mathbf{CM}}=-\frac{1}{2}\frac{\hslash^2}{2m}\nabla^2_{r_{\mathbf{CM}}}+m\omega^2\mathbf{r}_{\mathbf{CM}}^2
\end{equation}
\mbox{}
\noindent and relative coordinates:
\begin{equation}
\mathcal{H}_{0,\mathbf{rel}}=-2\frac{\hslash^2}{2m}\nabla^2_{r_{\mathbf{rel}}}+\frac{1}{4}m\omega^2\mathbf{r}_{\mathbf{rel}}^2
\end{equation}
respectively, according to the coordinate transformations 
${\mathbf{r}_{cm}=\frac{\mathbf{r}_1+\mathbf{r}_2}{2}}$ and $\mathbf{r}_{rel}=\mathbf{r}_1-\mathbf{r}_2$. 
\mbox{}\\

Thus, the total wavefunction attains the form:
\begin{equation}
\Psi=\Psi(\mathbf{r}_{\mathbf{CM}})\Psi(\mathbf{r}_{\mathbf{rel}})
\end{equation}

Since the centre-of-mass is trivially decoupled from the relative coordinates (inducing eventually a shift to the total energy of the system) we focus in \ref{relwf} on solving the relative coordinates stationary problem: $\mathcal{H}_{\mathbf{rel}}\Psi(\mathbf{r}_{\mathbf{rel}})=E\Psi(\mathbf{r}_{\mathbf{rel}})$. In fact, $\mathcal{H}_{\mathbf{rel}}$ should include also the interaction potential, however, due to its hard core form, it can be taken into account as a boundary condition on the harmonic potential solutions, as explained in \ref{2bcondition}.

To simplify the mathematical notation, from here on, we will consider dimensionless quantities, obtained through scaling of energies with $\hbar \omega$ and lengths with the characteristic oscillator length $\alpha_{osc}=\sqrt{\frac{\hbar}{m \omega}}$.

\subsection{The relative wavefunction}\label{relwf}

The solution for the Schr\"{o}dinger equation of the relative part $H_{rel}$ can be written as a product of a radial and an angular term: 

\begin{equation}
\Psi(\mathbf{r})=R(r)\cdot e^{il\phi},~~~~ l=0,1,2,...
\end{equation}

\noindent This ansatz results to the following equation for the radial part:

\begin{equation}\label{eq:318}
\Bigg(-\frac{1}{r}\frac{\partial}{\partial r}\Big(r\frac{\partial}{\partial r}\Big) + \frac{l^2}{r^2} + \frac{1}{4} r^2\Bigg)\cdot R(r)=E\cdot R(r)
\end{equation}

\noindent After a few algebraic steps, the solution of Eq.~(\ref{eq:318}) reads:
\begin{equation}\label{eq:339}
R(r)=r^l e^{-\frac{1}{4}r^2}F(r)
\end{equation}
where the function $F(r)$ is a solution of the confluent hypergeometric differential equation (see Appendix A). In particular, $F(r)$ is the linear combination of the first and second kind Kummer's functions:
\begin{equation}\label{eq:maiden}
\begin{split}
F(r)&= A\cdot M\bigg(\frac{1}{2}\Big(l+1-E \Big) ,~~ l+1,~~\frac{1}{2}r^2\bigg) +\\&+ B\cdot U\bigg(\frac{1}{2}\Big(l+1-E \Big)  ,~~ l+1,~~\frac{1}{2} r^2\bigg)
\end{split}
\end{equation}

\noindent Since $l+1$ is a positive integer, U becomes: 

\begin{equation}\label{eq:337}
\begin{split}
&U(a,l+1,\frac{1}{2}r^2)=\frac{(-1)^{l+1}}{l!\Gamma(a-l)}\bigg[ M(a,l+1,\frac{1}{2}r^2)ln(\frac{1}{2}r^2) +\\[10pt] & +\sum_{k=0}^{\infty}\frac{(a)_k(\frac{1}{2}r^2)^k}{(l+1)_kk!}\big(\psi(a+k)-\psi(1+k)-\psi(1+l+k)\big)\bigg]+ \\[10pt] &+ \frac{(l-1)!}{\Gamma(a)}(\frac{1}{2}r^2)^{-l}M(a-l,1-l,\frac{1}{2}r^2)
\end{split}
\end{equation}
where $\psi(x)=\Gamma'(x)/\Gamma(x)$. The last term in Eq.~(\ref{eq:337}) is taken to be zero for $l=0$.\\

Since the two particles are bound due to the harmonic potential, $R(r)$ should vanish for $r\rightarrow \infty$.
On the one hand, the term $$e^{-\frac{1}{4}r^2}M\bigg(\frac{1}{2}\Big(l+1-E \Big),~~ l+1,~\frac{1}{2} r^2\bigg)$$ satisfies the appropriate asymptotic condition only if ${\frac{1}{2}\Big(l+1-E \Big)=-N}$, where $N$ is a positive integer. 
On the other hand, the term $$e^{-\frac{1}{4}r^2}U\bigg(\frac{1}{2}\Big(l+1-E \Big),~~ l+1,~\frac{1}{2}  r^2\bigg)$$ satisfies the condition for ${\frac{1}{2}\Big(l+1-E \Big)=-m}$, where $m$ is a positive real number. When $m$ becomes a positive integer $U\propto M$. Therefore, without loss of generality we can keep only $U$ in Eq.~(\ref{eq:maiden}) and set $A=0$.

\subsection{Hard core interaction condition and energy spectrum} \label{2bcondition}

The energy eigenvalues of the interacting system are determined through the Dirichlet boundary condition at $r=r_0$:

\begin{equation}\label{eqcodnition}
R(r) \Big|_{r=r_0}=0~ \Rightarrow~ U\Big(-m,~l+1,~\frac{1}{2}r_0^2 \Big)=0
\end{equation}
implied by the form of the hard-core interaction potential in Eq.~(\ref{eq:2}). We solve numerically Eq.~(\ref{eqcodnition}) for fixed values of $l$ and $r_0$ to obtain a set of eigenenergies $SE_{l,r_0}$ characterized by different values of $m$ (the first argument of function U),  according to the expression:
\begin{equation}
\label{eqmfinder}
E_{SE_{l,r_0}}=l+1+2m
\end{equation}
We will refer in the following to the term "m-order" ($m_o$,  $o=1,2,...$ where $m_o<m_{o+1}$), which denotes the ascending values of $m$ within a specific set $SE_{l,r_0}$. Clearly, with this classification the entire spectrum (for given $r_0$) is obtained as $\displaystyle{\bigcup_{l=0}^{\infty}} SE_{l,r_0}$.

\section{ENERGY SPECTRUM ANALYSIS}
In this section we exploit the solution we have found in the previous section and investigate first the energy spectrum in subsection \ref{levelcross} which we analyse in the following subsections in terms of potential and kinetic energy contributions (\ref{potkin}), wave profile (\ref{wfprof}) and Fisher Information (\ref{Fisherinfo}). 

\subsection{Energy spectrum bunching and level crossings}\label{levelcross}
\begin{center}
\begin{figure}[H]
\centering
\includegraphics[scale=0.25]{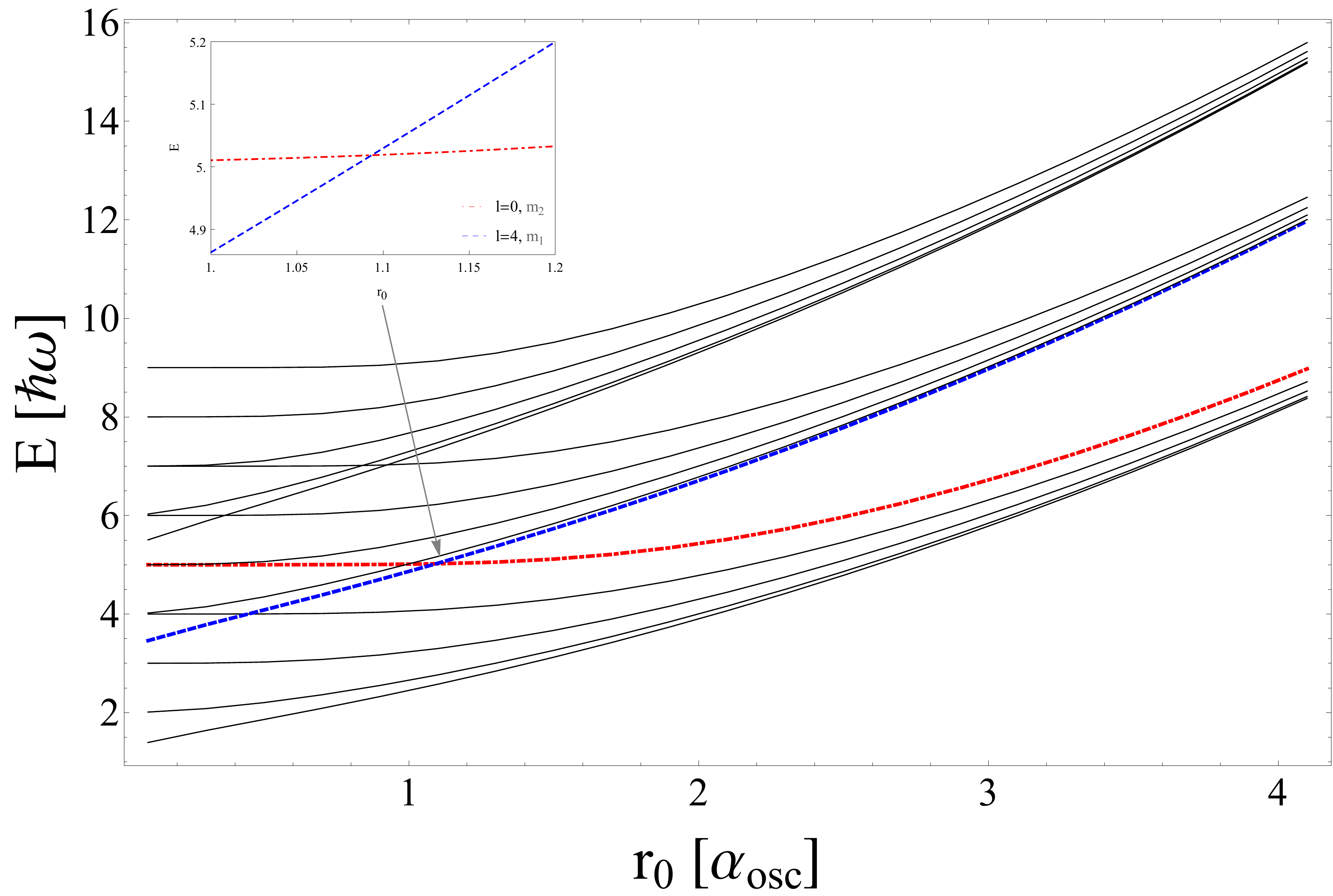}
\caption{(Color online) The energy spectrum of two hard-core bosons in a two-dimensional harmonic trap as a function of the range of the hard-core interaction $r_0$. For small $r_0$ we approach the non-interacting spectrum while for $r_0 \gg 1$ a gradual bunching of states occurs. The inset focuses on the level crossing of two particular energy levels, ($l=0,m_2$)(blue dashed line) and ($l=4,m_1$) (red dotted-dashed line). The level crossing is general, occurring for several energy eigenstates of the system. For all these crossings the underlying mechanism is common. Thus, for its explanation it suffices to focus on the particular states presented in this figure.}
\label{fig1}
\end{figure}
\end{center}

In Fig.~\ref{fig1} we show the energy spectrum of the system as a function of $r_0$. We observe that for $r_0\ll 1$, the spectrum approaches that of the non-interacting 2D harmonic oscillator, where the energy levels are given by the expression $\epsilon_{\text{free}}=l+1+2\tilde{m}$, for $\tilde{m}$ positive integer with the corresponding degeneracy due to the symmetry of the potential. Therefore we refer to $r_0\ll 1$ as the low-correlation regime.

In the opposite limit, $r_0\gg 1$, there is a bunching of states, corresponding to the same $m$-order, into bands. More explicitly these states are coming from different sets $SE_{l,r0}$ with $m$ of the same m-order $m_o$ within each set.
For $r_0 \to \infty$ the bunching tends to a continuous band. The reason is that for two levels with consecutive $l$ the energy difference $\Delta E = \Delta l + 2 \Delta m$ with $\Delta l =1$, $-0.5< \Delta m<0$ shrinks with increasing $r_0$ since $\Delta m$ is a decreasing function of $r_0$. Furthermore, for every level, $m$ increases rapidly with $r_0$ and becomes the dominating term in Eq.~(\ref{eqmfinder}) when $r_0\gg 1$. As a result, we see a similarly rapid increase of each energy level $E_{l,m}$. In fact, one can consider the formation of the band-like structure as the two-dimensional imprint of  fermionization, lacking, however, a direct mapping to a strict "fermionic" limit as in the 1-D case. 

The most prominent observation for the energy spectrum, lies in the intermediate regime between those two limits where several energy level crossings occur. This phenomenon takes place \textbf{only} between levels of different $m$-order from different sets $SE_{l,r0}$. A highlighted such crossing, on which we will focus to understand the general mechanism, is that between the ($l=0,m_2$) and the ($l=4,m_1$) level shown in the inset of Fig.~\ref{fig1}. Slightly before the crossing point, those are the 5th and 6th energy level respectively and interchange their order right after that point.

\subsection{Potential and kinetic energy}\label{potkin}

A first step, towards understanding the level-crossing behaviour, is done by separately calculating the mean kinetic  
\begin{equation}
\int_{r_0}^{\infty} \Psi^{*} \frac{1}{4}r^2 \Psi r dr
\end{equation}
and the mean potential energy:
\begin{equation}
\int_{r_0}^{\infty} \Psi^{*} \big(-\nabla_{r}^2 +\frac{l^2}{r^2}  \big)\Psi r dr
\end{equation}
for a state $\Psi$.

\begin{figure}[H]
\centering
\includegraphics[scale=0.25]{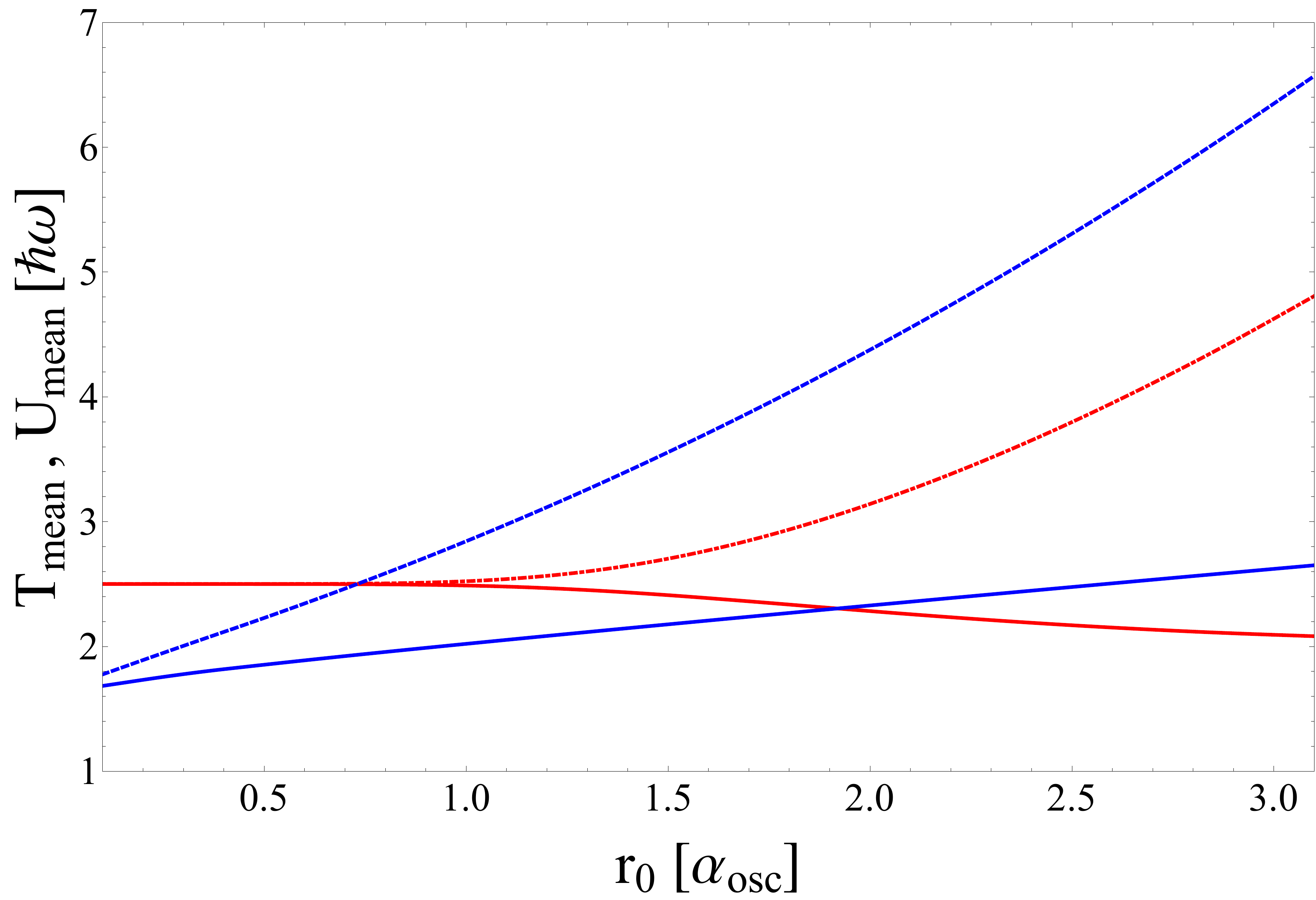}
\caption{(Color online) Mean kinetic and potential energy for the states $(l=0,m_2)$ (blue and blue dashed line respectively)  $(l=4,m_1)$ (red and red dashed-dotted line respectively) as a function of the range of the hard-core interaction $r_0$.} 
\label{fig2}
\end{figure}

As shown in Fig.~\ref{fig2}, for the states $(l=4,m_1)$  and $(l=0,m_2)$, the kinetic energies remain relatively flat compared to the corresponding potential energies. Moreover, the rate of change for the potential energy of $(l=0,m_2)$ with $r_0$ is much greater than that of $(l=4,m_1)$ which results to the corresponding crossing. This mechanism is general (as we show in Appendix B) and has its origin on the behaviour of the corresponding wavefunctions as $r_0$ increases, as we will explain in the next section. 

\subsection{Wavefunction profile}\label{wfprof}

\begin{figure}[H]
\centering
\includegraphics[scale=0.25]{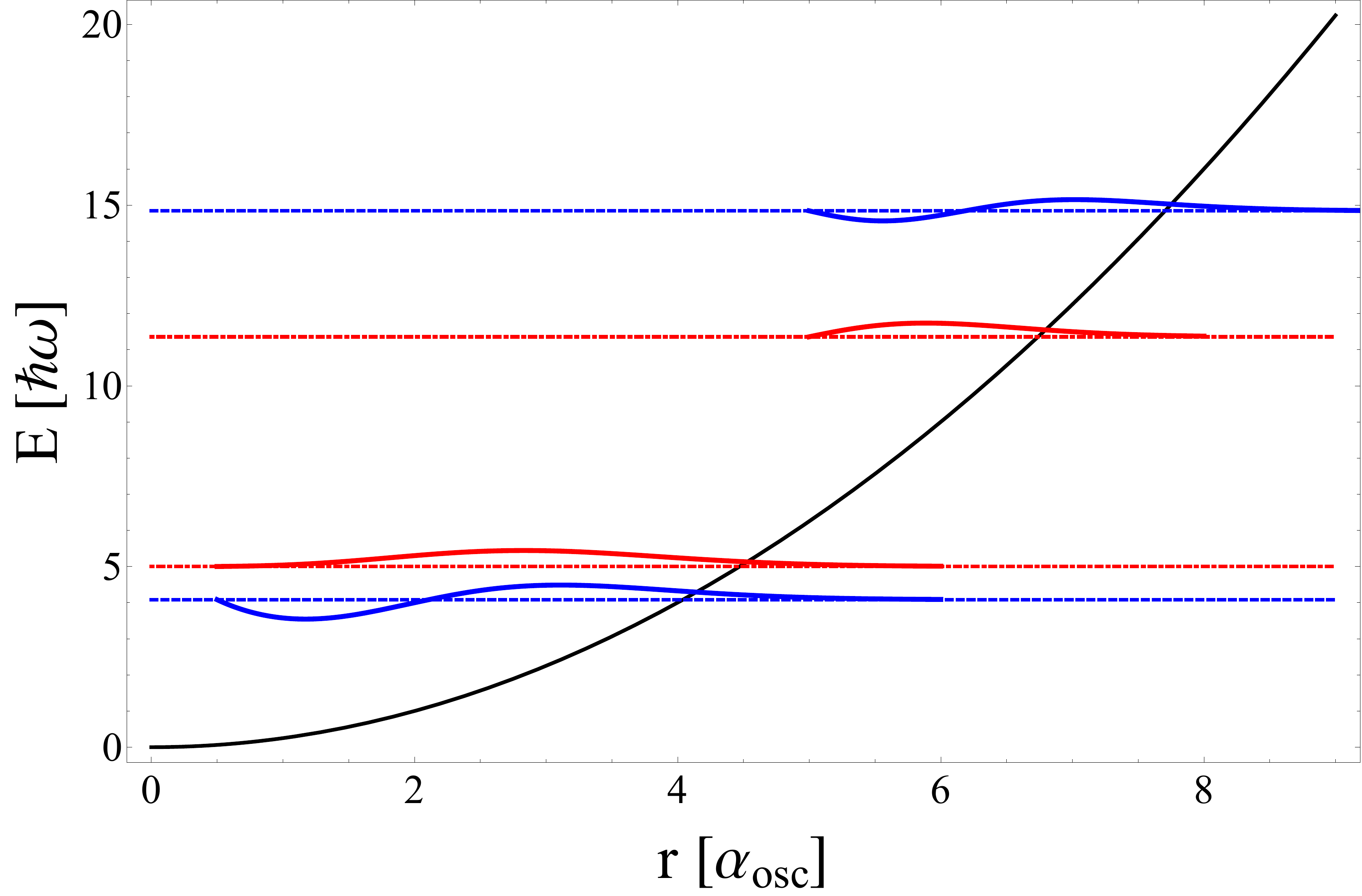}
\caption{(Color online) The radial part of the wavefunction $R(r)$ for the states $(l=0,m_2)$ (blue,dashed) and  $(l=4,m_1)$ (red, dot-dashed), plotted for two values of the hard-core range $r_0$, before and after the crossing point (lower layer and upper layer respectively). The shift to the right is leading to a compression of both, which is more abrupt for  $(l=0,m_2)$.}
\label{fig3}
\end{figure}

A next step towards the understanding of the structure of the energy spectrum and particularly the crossing mechanism, is achieved by considering the modulation of the corresponding relative-coordinate wavefunctions with increasing range $r_0$ of the hard-core interaction potential.
First, let us mention that the number of nodes of each wavefunction corresponds to the order of $m$ from the solutions of Eq.~(\ref{eqmfinder}). In Fig.~\ref{fig3} we show how, increasing the hard-core range $r_0$, the wavefunction profile shifts to the right (larger $r$ values) and becomes at the same time more "compressed". 

In fact, the  $(l=0,m_2)$ state, having two nodes, becomes much more modified and compressed than the $(l=4,m_1)$ state. As a consequence, the potential energy of the former is more sensitive to the change of $r_0$. Since the two states differ in the number of nodes and therefore in their oscillatory behaviour, the state with lower number of nodes displays less variability with $r_0$. Stated in a different way, the nodes form a channel for quantum correlations induced to the wavefunction through the $r_0$ variation. To illustrate further this issue, we will employ the Quantum Fisher Information (QFI) as a tool to quantify these correlations in the next section.

\subsection{ Quantum fisher information}\label{Fisherinfo}

We have understood in the previous section qualitatively the reason why the wavefunction of $(l=0,m_2)$ is more sensitive to the change of $r_0$ and the resulting larger rate of potential energy change that leads to the crossing with ($l=4,m_1$). 
In order to give a quantitative measure for the local behaviour of the probability density, as well, as its sensitivity to the variation of the hard-core range of the interaction $r_0$, we calculate the QFI for the crossing states. 

\begin{center}
\begin{figure}[H]
\centering
\includegraphics[scale=0.28]{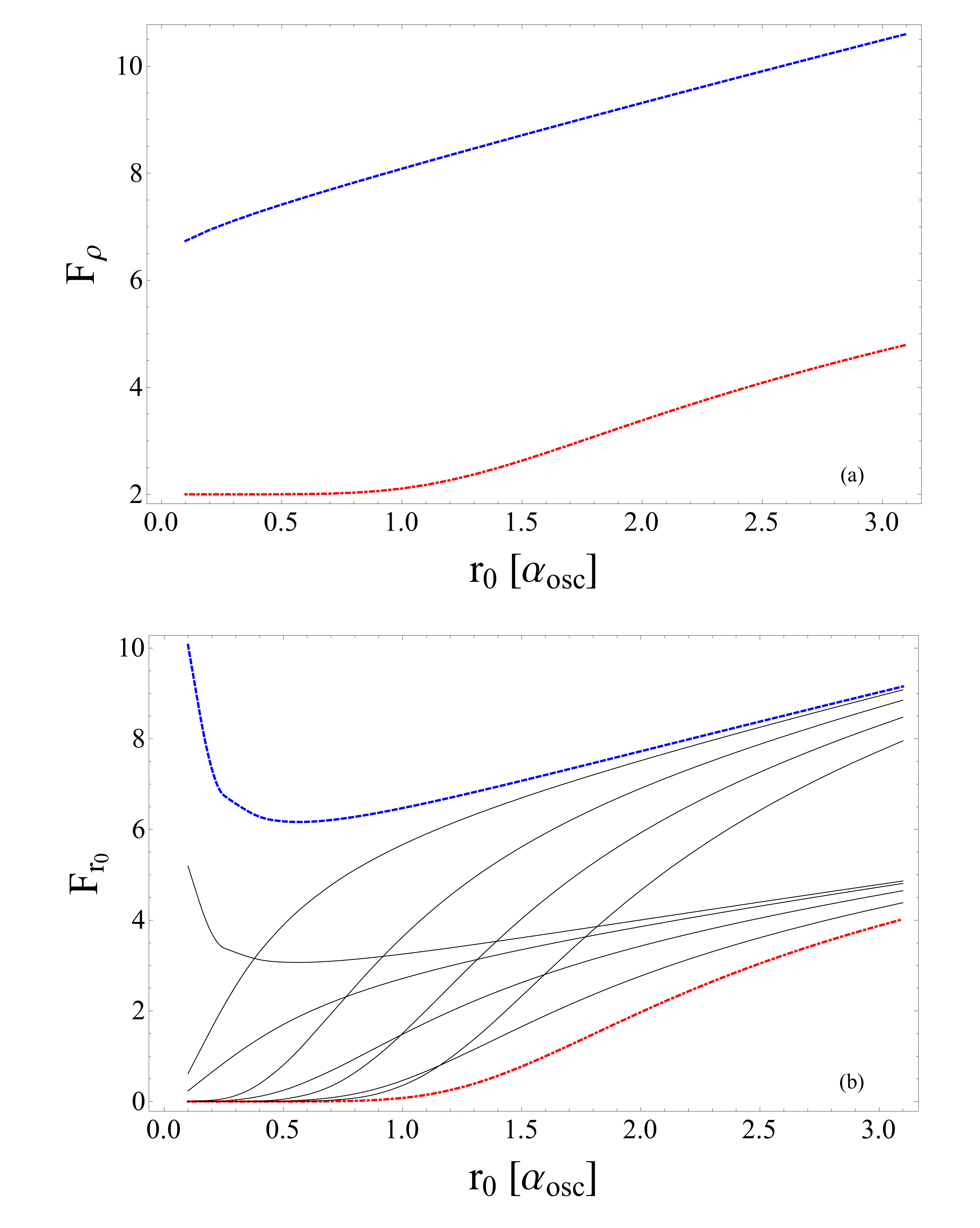}
\caption{(Colour online) QFI as a function of the hard-core range $r_0$, calculated with two different methods based on (a) density $\rho$, and (b) $r_0$ as explained in the text. For the two considered states $(l=0,m_2)$ (blue dashed) and  $(l=4,m_1)$ (red dot-dashed), QFI shows either the difference in the localization and oscillatory behaviour of the density (a), or the sensitivity to $r_0$ (b).}
\label{fig4}
\end{figure}
\end{center}

There are at least two possibilities to define and calculate QFI in our system depending on the issue one wishes to resolve. The first one focuses on the properties of the density profile. In this context, QFI is given in terms of $\rho$ by \cite{romera}:

\begin{equation}
F_{\rho}=\int_{\mathcal{R}^3}\frac{\abs{{\nabla}\rho(\mathbf{r})}^2}{\rho(\mathbf{r})} d^3r = \int_{\mathcal{R}}\frac{\abs{{\nabla}\rho_{rel}(r)}^2}{\rho_{rel}(r)} rdr 
\end{equation}

In Fig.~\ref{fig4}(a) we observe that $F_{\rho}$ is always larger for $(l=0,m_2)$. This quantifies the degree of "localization" of the density around its nodes as well as the number of those nodes. Therefore $F_{\rho}$ helps us to explore quantitatively which states have a higher degree of oscillatory behaviour (more nodes and localization around them). Obviously,  the observed QFI behaviour supports the argument that the wavefunction nodes provide a channel for increasing quantum correlations through a parameter change.

Due to this oscillatory behaviour and corresponding compression and localization, it is expected that the state $(l=0,m_2)$ is more sensitive to the change of $r_0$. This is quantitatively accessible by the second approach of calculating QFI for pure states  \cite{kinezoikaluteroi} as a function of a parameter (here $r_0$) which reads:

\begin{equation}
F_{r_0}=4\Big(\bra{\partial_{r_0}\mathcal{\psi}}\ket{\partial_{r_0}\mathcal{\psi}}-\abs{\bra{\mathcal{\psi}}\ket{\partial_{r_0}\mathcal{\psi}} }^2          \Big)
\end{equation}

In Fig.~\ref{fig4}(b) we can see the behaviour of $F_{r_0}$ for several states but also for the two particular states that are on focus in our investigation. One can clearly see that $(l=0,m_2)$ has always a larger value of $F_{r_0}$ than $(l=4,m_1)$ and also changes more abruptly as $r_0$ increases, both indicating that the sensitivity of this state to the change of the hard-core range, is greater.

\begin{center}
\begin{figure}[H]
\centering
\includegraphics[scale=0.25]{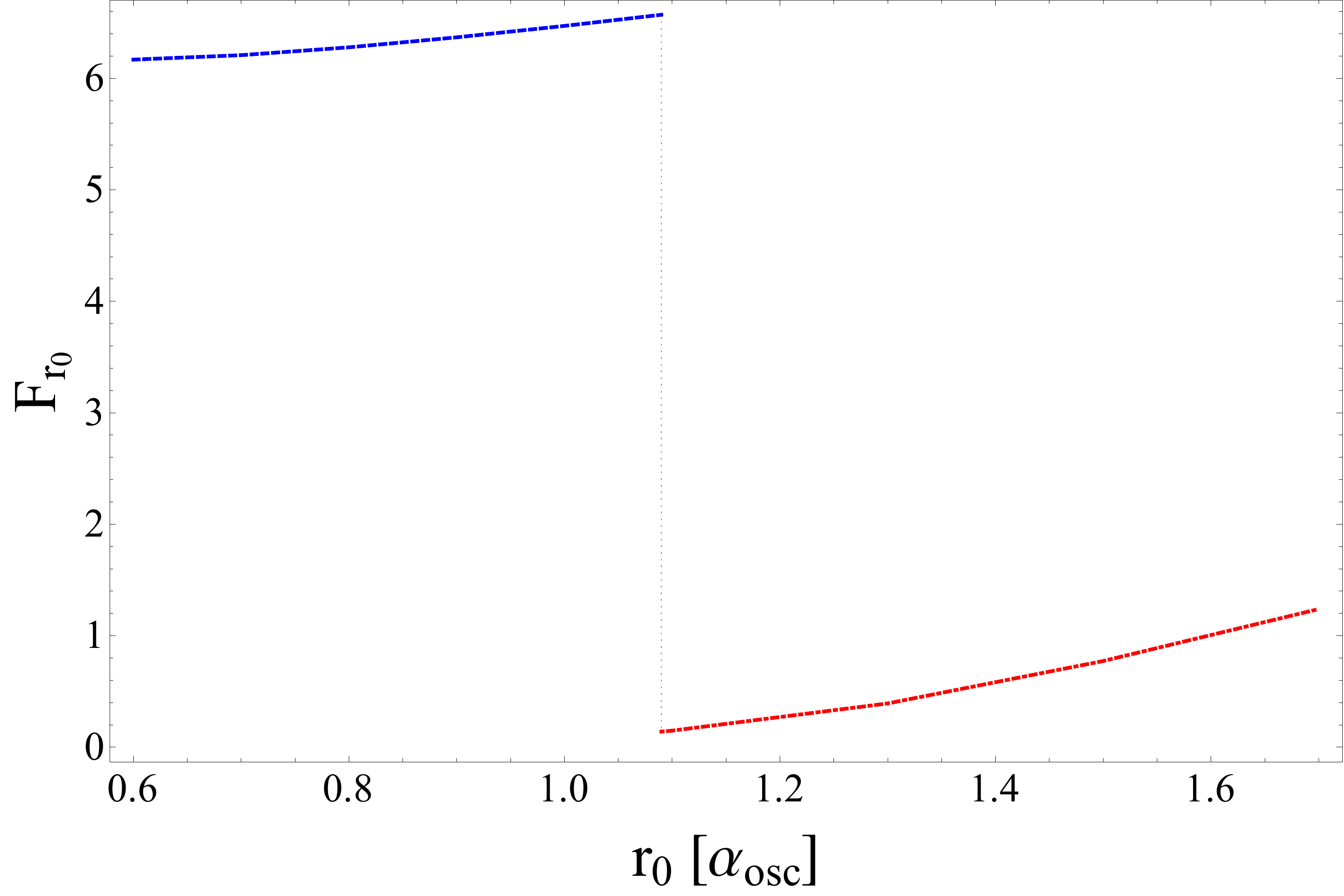}
\caption{(Colour online) Discontinuity of QFI as a function of $r_0$ that depicts the change of the eigenfunction that corresponds to the fifth energy level where $(l=0,m_2)$ (blue dashed) and $(l=4,m_1)$ (red dot-dashed).}
\label{fig7}
\end{figure}
\end{center}

As a last step, and adopting the viewpoint represented in \cite{levelcrossing_fisher}, we consider $F_{r_0}$ not for a certain state but for the two levels that are crossing. In particular we take level 5 before the crossing that will inverse their order after the crossing. This will come with an abrupt change, more specifically a discontinuity or a "gap" in QFI, corresponding to a large change on entanglement properties, as we depict in Fig.~\ref{fig7}

\section{CONCLUSIONS AND OUTLOOK}

We performed an extensive study of the spectral properties of a system consisting of two identical hard disc spinless bosons trapped in a two dimensional isotropic harmonic trap. An analytical form for the corresponding wavefunctions in terms of hypergeometric functions has been obtained. The crucial dimensionless parameter determining the energy spectrum of the system is the ratio of the hard-core potential radius to the characteristic length scale of the harmonic trap. For small values of this ratio the system's energy spectrum is characterized by multiple level crossings, while when this ratio increases, a level bunching is formed. We have been able to explain this behaviour in terms of the difference in the variation of the potential energy part, when tuning the hard-core potential radius for two exemplary levels, participating in a crossing. This difference is demonstrated also in the corresponding wavefunctions which turn out to posses different number of nodes and localization/oscillatory behaviour between these nodes. Furthermore, we have calculated the Quantum Fisher Information for crossing levels in two different ways, taking as relevant parameter (i) the associated probability density and (ii) the radius of the hard-core potential. We have shown that the QFI increases for increasing degree of localization around the wavefunction nodes and displays a discontinuity as a function of the hard-core radius when passing through the level crossing point. Our analysis reveals the role of localization between the wave function nodes as an alternative source of quantum correlations different than entanglement. This may be an essential property of quantum many body systems eventually associated with the creation of quantum multi-particle correlations. Investigating this issue in detail is certainly an interesting task which goes beyond the scope of the present work.

\section*{ACKNOWLEDGMENTS}
I. B. acknowledges financial support by Greece and the European
Union (European Social Fund - ESF) through the Operational
Programme “Human Resources Development, Education
and Lifelong Learning” in the context of the project
“Reinforcement of Postdoctoral Researchers" (MIS-5001552),
implemented by the State Scholarships Foundation (IKY).

\section*{APPENDIX A: DERIVATION OF CONFLUENT HYPERGEOMETRIC SOLUTIONS } \label{AppendixA}

To simplify Eq.~(\ref{eq:318}) we introduce:

\begin{equation}
u(r)=R(r) \cdot r^{\frac{1}{2}} 
\end{equation}

Hence Eq.~(\ref{eq:318}) becomes:
\begin{equation}
\Bigg(-\frac{d^2}{dr^2}+\frac{l^2-\frac{1}{4}}{r^2}+\frac{1}{4}r^2\Bigg)\cdot u(r)=E\cdot u(r)
\end{equation}
\mbox{}

Setting $b=l^2-\frac{1}{4}$, the above equation takes the form:

\begin{equation}
\Bigg(\frac{d^2}{dr^2}-\frac{1}{4} r^2-\frac{b}{r^2}+E\Bigg)\cdot u(r)=0
\end{equation}

Via the transformation:

\begin{equation}\label{bbb}
u(r)=r^ce^{\kappa r^2}F(r)
\end{equation}

\noindent where $c$ and $\kappa$ are unknown constants, the Differential Equation (DE) will eventually end up to the Confluent Hypergeometric DE. 

Imposing the transformation we have:

\begin{equation}\label{eq:aaa}
\begin{split}
&\frac{d^2F}{dr^2}+\bigg(2c\frac{1}{r}+4\kappa r\bigg)\frac{dF}{dr}+\\&+\bigg[\big(c(c-1)-b\big)\frac{1}{r^2}+(4\kappa^2-\frac{1}{4})r^2+2(2c+1)\kappa+E  \bigg]\cdot F=\\&=0
\end{split}
\end{equation}

At this point, we introduce the transformation:
\begin{equation}
z=\lambda r^2
\end{equation}
\noindent where $\lambda$ is a real constant.

So Eq.~(\ref{eq:aaa}) can be written as:

\begin{equation}\label{iron}
\begin{split}
&4\lambda z\frac{d^2F}{dz^2}+\bigg(2\lambda(1+2c)+8\kappa
z\bigg)\frac{dF}{dz}+\\&+\bigg[\big(c(c-1)-b\big)\frac{\lambda}{z}+(4\kappa^2-\frac{1}{4})\frac{z}{\lambda}+2(2c+1)\kappa+E  \bigg]\cdot F=\\&=0
\end{split}
\end{equation}
\noindent with respect to $z$.

Eq.~(\ref{iron}) can be mapped to the Confluent Hypergeometric DE, by setting constants $c$ and $\kappa$ :

\begin{equation}\label{ccc}
b=c~(c-1)
\end{equation}
\begin{equation}
\kappa^2=\frac{1}{16}
\end{equation}

Solution (\ref{bbb}) needs to vanish for $r\rightarrow\infty$, so we keep the negative value of $\kappa$:

\begin{equation}
\kappa =-\frac{1}{4}
\end{equation}
Considering the definition of $b$ as well as Eq.~(\ref{ccc}) we have:

\begin{equation}
c~(c-1)=l^2-\frac{1}{4}
\end{equation}

that is fulfilled for:
\begin{equation}
c=l+1
\end{equation}

Having specified above the values of the constants $c$ and $\kappa$, we set $\lambda=\frac{1}{2}$ resulting to the Confluent Hypergeometric DE:

\begin{equation}
z\frac{d^2F}{dz^2}+\Big(l+1-z\Big)\frac{dF}{dz}-\frac{1}{2}\Big(l+1-E \Big)  \cdot F=0
\end{equation}

\section*{APPENDIX B: GENERAL MECHANISM FOR LEVEL CROSSING} \label{AppendixB}
\begin{center}
\begin{figure}[]
\centering
\includegraphics[scale=0.25]{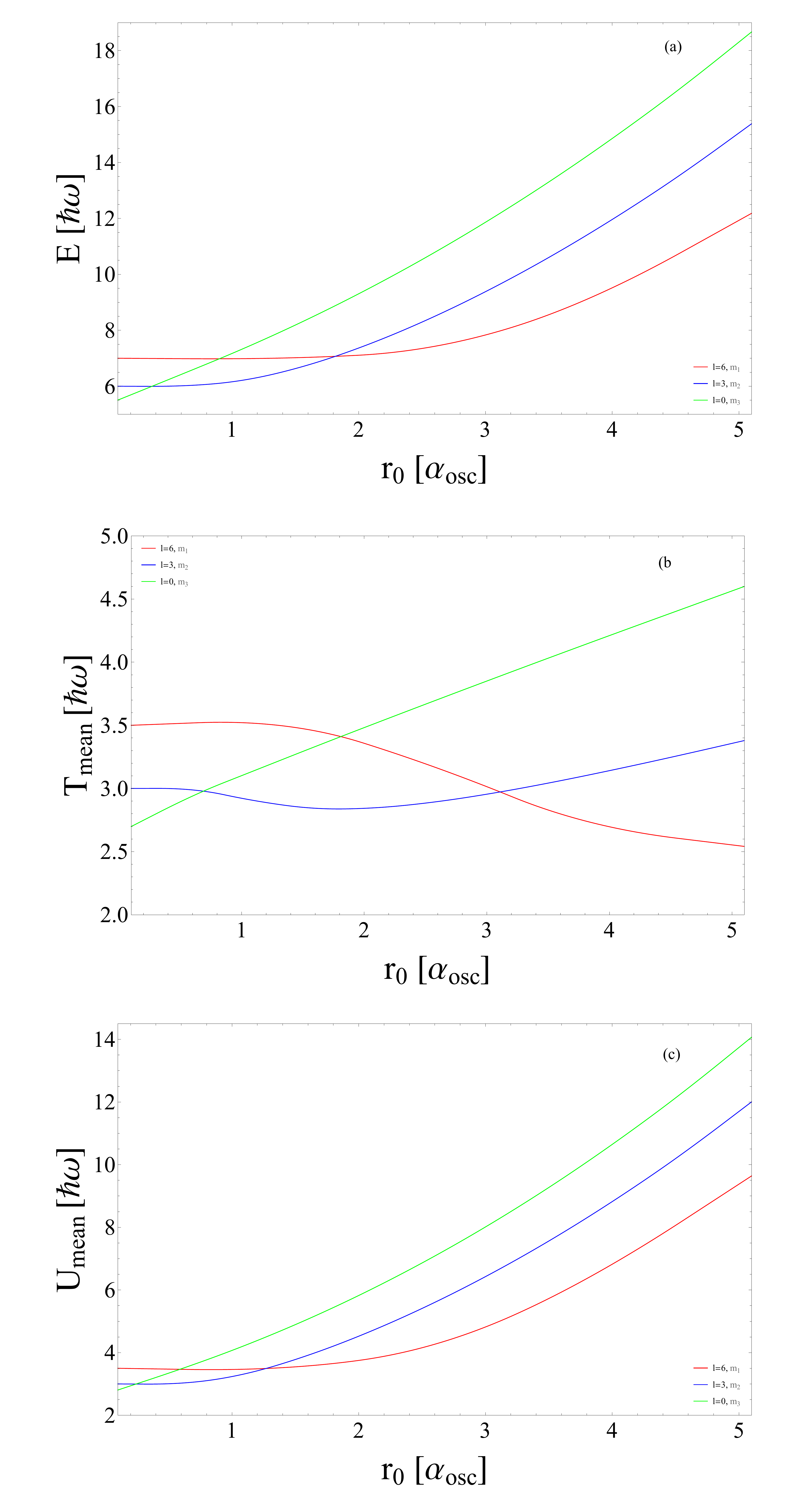}
\caption{(a) Level $(l=6,m_1)$ (red) having two crossing points with levels $(l=0,m_3)$ (green) and $(l=3,m_2)$ (blue) by varying $r_0$. The panels (b), (c) demonstrate behavioural similarities with Fig.~\ref{fig2} in terms of mean kinetic and potential energies as functions of $r_0$, indicating the generality of the proposed level-crossing mechanism.}
\label{fig6}
\end{figure}
\end{center}

This section presents the level crossing scenario in the more general case when an energy level (here $l=3,m_2$) crosses with two other levels, one with higher ($l=0,m_3$) and one with lower ($l=6,m_1$) order. The observed behaviour is summarized in Fig.~\ref{fig6} where, for completeness, we also show the variations of the mean kinetic and potential energy as a function of $r_0$, to reveal the generality of the level crossing mechanism described in subsection \ref{potkin}. Notice that the behaviour in the immediate neighbourhood of each crossing point is similar to the one examined 
in Fig.~\ref{fig2}.

\end{document}